# Rational Design Principles for Na- and Li-ion Carbon Anodes from Interlayer Spacing Control

Ihor RADCHENKO[1], Oleksandr I. MALYI[1,2,*]

[1]Centre of Excellence ENSEMBLE3 Sp. z o. o., Wolczynska Str. 133, 01-919, Warsaw, Poland
[2]Qingyuan Innovation Laboratory, 1 Xueyuan Road, Quanzhou 362801, P. R. China
[*]Corresponding author; email: oleksandrmalyi@gmail.com

**Abstract:** Graphite, the standard commercial anode for Li-ion batteries, is thermodynamically incompatible with Na-ion batteries, leading researchers to search for alternative C-based structures (e.g., hard carbon, expanded graphite). In a simplified picture, the main idea of such search relies on identifying disordered C structures with a large interlayer spacing and distribution of local structural motifs (e.g., pores) with target electrochemical properties. Such exploration is typically done via trial-and-error experimentation, which often does not allow precise understanding of the role of interlayer distance and even Na/Li-ion intercalation in the electrochemical performance. Motivated by this, using density-functional theory and cluster expansion, we establish a structure-property relationship for Li- and Na-intercalation across a range of graphite interlayer spacings and stacking arrangements. We show that Na intercalation becomes thermodynamically possible in large concentrations above 4.21 Å even without a change in interlayer spacing. Conversely, Li intercalation has a narrow optimal window, with maximum capacity (close to the commercial limit) at approximately 3.75 Å, while larger spacings (e.g., 4.58 Å) quickly reduce Li storage capacity. We also find that AA-stacked domains consistently offer stronger ion bonding and higher voltages than AB-stacked domains for both ions. Our results, thus, not only explain the role of metal ion intercalation in the electrochemical performance, but also clarify the fundamental design trade-offs for expanded C anodes, offering practical targets and interlayer distance ranges for the independent optimization of the next generation of negative electrodes for metal-ion batteries.

**Keywords:** d-spacing, Li/Na intercalation, DFT, cluster expansion, expanded graphite, hard carbon.

# 1. Introduction

Materials design sets the limits of rechargeable batteries, with structure and thermodynamics determining battery operational voltage and electrode material performance [1]. While graphite is a commonly used commercial electrode material for Li-ion batteries due to its high capacity of 372 mAh g$^{-1}$ and low volume expansion [2–4], its application for Na-ion batteries is limited. The fundamental reason for such behavior is thermodynamics. Specifically, in contrast to Li, which can strongly interact with graphite and form a stable $LiC_6$ compound, Na intercalation into graphite is thermodynamically unfavorable [5]. The inability of Na to intercalate into the tightly spaced carbon layers of graphite has driven researchers to explore alternative carbon anodes that feature more complex atomistic structures, like expanded graphite [6] or hard carbon [7] for Na-ion batteries. The main idea behind their design lies in (i) increasing the interlayer distance in the graphite-like domains, (ii) introducing the higher energy sites (e.g., point defects and non-hexagonal ring defects), and (iii) engineering internal closed nanopores that serve as reservoirs to store active ions. In this way, Na atoms have more reactive sites for chemical bonding, which minimizes the strain caused by their large size, leading to more favourable binding energy. These structural motifs not only benefit Na storage but can also be efficiently utilized in Li-ion batteries. For instance, hard carbon, considered the most commercially attractive material for Na-ion storage (due to its preparation simplicity and high reversible capacity, which can routinely be larger than 300 mAh/g [8–12]), can also be utilized for Li-ion batteries, with a reversible capacity even exceeding that of graphite (>400 mAh/g) [13,14]. An alternative carbon architecture that follows the same design logic is expanded graphite. Similar to hard carbon, it can exhibit superior Na storage behavior [6] and competitive Li-ion electrochemical performance [15,16] when defect density, turbostratic disorder, and surface chemistry are tuned appropriately. Taken together, the behaviour of hard carbon and expanded graphite underscores that increased interlayer distance, defects, and other structural motifs play a decisive role in governing Na- and Li-ion storage, but the exact contribution of each factor is still under debate. Both hard carbon and expanded graphite possess very complex atomic structures, combining amorphous regions, graphite-like crystalline domains (with multiple possible stacking orders), and various structural and chemical defects. This complex composite structure makes it challenging to attribute the experimentally observed storage performance to individual structural motifs. In particular, there are somewhat contradictory reports in the literature on the role of Na intercalation into graphite-like layers in the storage. While a large reversible capacity is realized in the carbon-based electrodes, some experimental studies suggest that Na intercalation does not occur in the crystalline domains of hard carbon (based mainly on apparent changes in interlayer spacing during battery operation) [17–19], while, in contrast, others discuss the possibility for Na storage in expanded graphite systems [6]. Motivated by this gap, herein, we develop an accurate and predictive description of Li and Na storage across the expanded graphite system, clarifying the interlayer distance thresholds at which Na intercalation becomes thermodynamically viable and to what extent. Following this analysis, we establish a set of structure–property relationships and practical targets (e.g., $d_{002}$ windows) that enable the rational design of carbon anodes for both Li- and Na-ion storage. Importantly, we also explain the differences in storage behavior for the two active ions, shedding light on why some materials that work well for Na-ion batteries may have somewhat limited performance in Li-ion applications.

## 2. Results and Discussion

While, from a theoretical perspective, AB graphite (space group: $P6_3/mmc$) is a thermodynamically stable form of carbon, in the context of C-based electrode materials one usually expects that AB graphite will coexist with AA graphite (space group: $P6/mmm$), which, according to state-of-the-art DFT calculations, have very comparable energetics and X-ray diffraction spectra [20,21]. This is not surprising, as both phases consist of $sp^2$-bonded graphene layers that interact weakly with each other via van der Waals (vdW) forces. What makes them distinct is the spatial arrangement of the graphene layers. Specifically, in AA graphite, the hollow sites of adjacent layers are aligned, whereas in AB graphite, the carbon atoms of one layer are aligned above the hollow sites of the adjacent layer (see Fig. S1 for an illustration). While this difference may appear insignificant, it actually affects the equilibrium interlayer distance and the Na-C interaction. For instance, DFT calculations using optB88-vdW exchange-correlation (XC) functional yield interlayer distances of 3.52 Å (AA graphite) and 3.33 Å (AB graphite) (see details in Table S1 for comparison of interlayer distances computed using different XC functionals). For pristine graphite, the Li intercalation energies in AA and AB graphite are -0.65 and -0.50 eV in the dilute limit (see Table S2). The energy difference arises because Na and Li prefer hollow sites on graphene [22]; AA-stacked graphite provides hollow coordination to both adjacent layers, whereas AB stacking leaves the ion hollow to one layer but at a top site relative to the other. To better understand Li and Na storage behaviours in graphite, we first focus on Li/Na insertion into AA- and AB-stacked graphite with equilibrium interlayer distance by combining first-principles cluster expansion with convex hull analysis (i.e., predicting which compounds are thermodynamically stable against any possible decomposition reaction at a given composition, see Fig. 1a-c). As expected, the results demonstrate that Li can be effectively stored in both AA and AB graphite via formation of a set of stable compounds (with respect to decomposition to competing phases), while this is obviously not the case for Na interaction with both AA and AB graphite. Our optB88-vdW convex hull calculations (using the respective reference state, like AA or AB graphite for the C side) reveal distinct stable compounds for different graphite stackings. For AA stacking, $LiC_{28}$, $LiC_{16}$, and $LiC_6$ are found to be stable, whereas for AB stacking, only $LiC_6$ forms (see Fig. 1b and Table S3). Consistent with single atom calculations, AA stacking typically provides stronger bonding; however, both stackings support the formation of stable $LiC_6$, a result also evident in the voltage profile shown in Fig. S2. We also find that Li and Na intercalation between graphite layers involves a significant expansion of the interlayer distance. For instance, for the most stable LiC$_6$ compounds, Li intercalation changes the interlayer distance from 3.52 to 3.62 Å (+2.8%) for AA-stacked graphite, and from 3.33 Å to 3.87 Å (+16.2%) for AB-stacked graphite (see Fig. 1b and Table S3). These results are in line with those reported in other works for Li-graphite-based systems [23–25]. In contrast to this comparatively modest and thermodynamically favourable lattice response for Li, Na insertion into graphite is accompanied by substantially larger, destabilising interlayer expansions. For Na, unstable NaC$_{28}$ compounds (with the lowest (positive) formation energy) increase the interlayer distances from 3.52 to 4.21 Å (+19.6%) for AA-stacked graphite and from 3.33 to 4.01 Å (+20.4%) (no Na present between layers) and 4.38 Å (+31.5%) (Na present) for AB-stacked graphite. It is important to note that the first-principles level description of the properties for a vdW system may not be straightforward, as, despite the advancement in the field, even the most utilized methods for describing vdW forces still have some error bars for the calculation [26]. Here, we find that the computed trends are generally robust, although the absolute formation energies can shift depending on the vdW functional (see Figs. S3-S6). A single exception is the PBE-TS approach that significantly overestimates cohesive energies for bulk Li and Na as compared to other functionals; for instance, the computed cohesive energy for bulk Na with PBE-TS is -2.57 eV, while the same value for optB88-vdW is -1.04 eV (Table S4). Also, for the case of the DFT-D2 approach, we find the possibility for the formation of compounds with higher Li content than LiC$_6$ (i.e., LiC$_3$, see Fig. S3) and even formation of NaC$_6$ compounds. Such large Li concentrations are not normally observed experimentally [4,27]. Similarly, Na storage in graphite system is known to be limited according to experimental results [28]. Taking this into account, we conclude that the formation heat of the Li-C and Na-C compounds in DFT-D2 calculations is somewhat overestimated compared to experimental results. In contrast, the cohesive energy predicted by vdW-DF and vdW-DF2 functionals (the first generation of the vdW XC functional aiming to account for vdW forces in terms of non-local correlation) is underestimated compared to experimentally known values (Table S4).

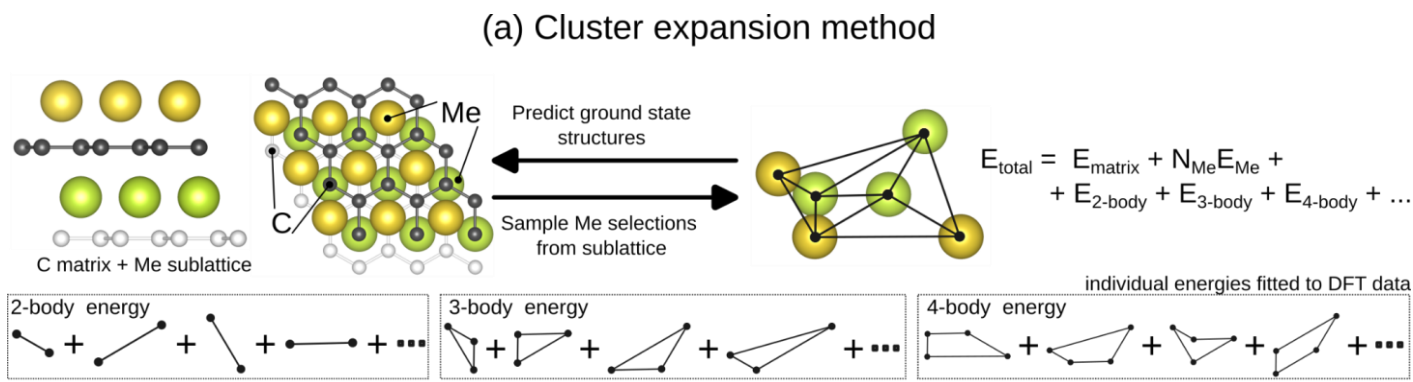
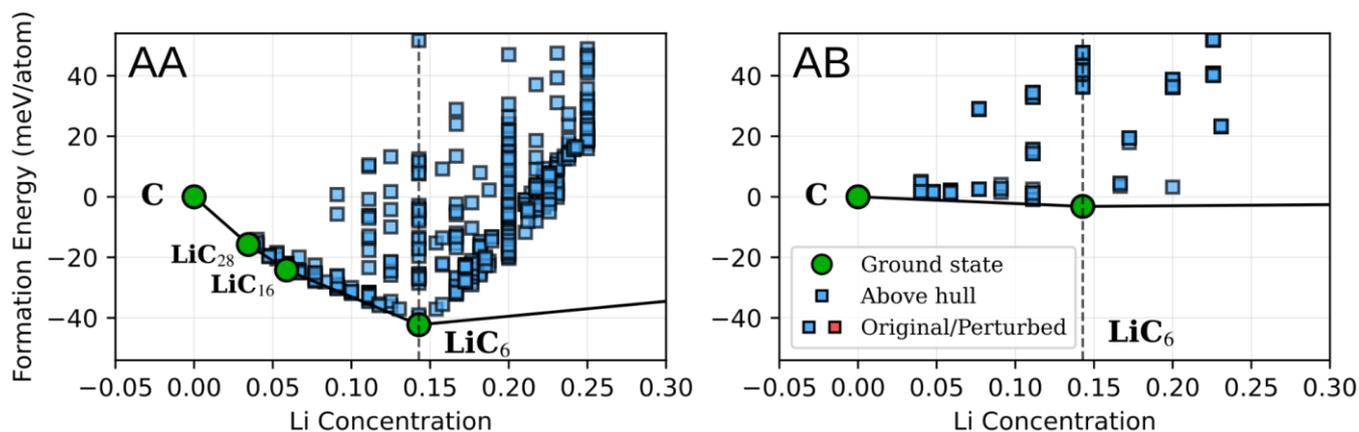
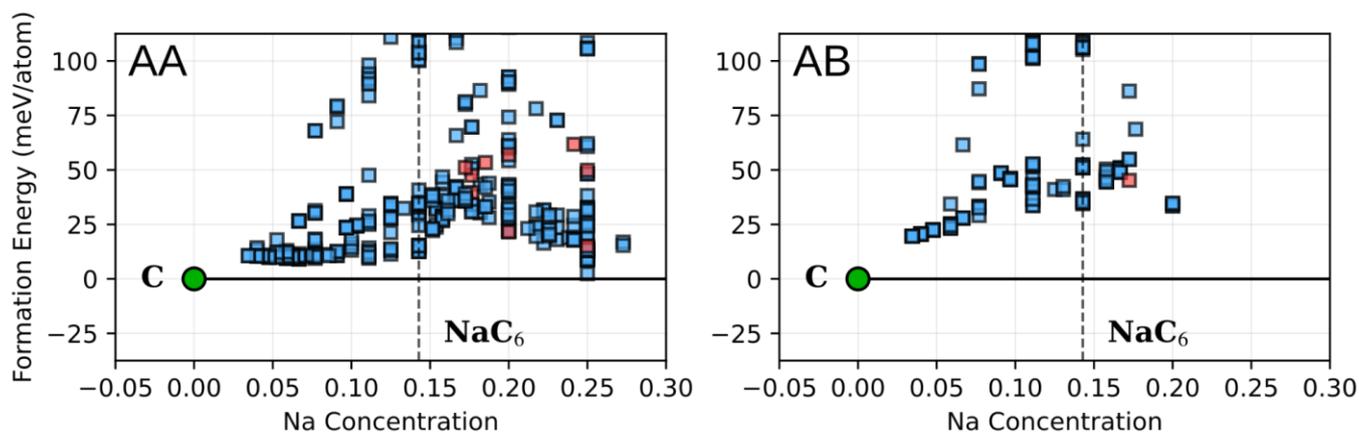

**Figure 1. Li and Na storage in unconstrained AA and AB graphite.** (a) Schematic of cluster expansion methodology for alkali-metal intercalation in carbon matrices. (b-c) Convex hull phase diagrams depicting formation energies as a function of Li/Na concentration for AA and AB stacking configurations in Li-C and Na-C systems. The formation energies are calculated relative to pure graphite with the same stacking and metallic Li/Na reference. All the results are given for the optB88-vdW exchange-correlation functional.

The above calculations represent the case of Li and Na storage in pure graphite, where the interlayer distance starts at the equilibrium and increases due to Li or Na intercalation. In the case of expanded graphite being used as an electrode material, the expanded nature of graphite is largely due to point defects/impurities stabilizing the large interlayer distance even in the absence of Na or Li [6,29]. Similarly, increased interlayer distance in graphite-like nanocrystalline domains inside hard carbon is stabilized by the complex interaction with amorphous regions and by the presence of other structural defects. Thus, Na and Li intercalation does not necessarily result in additional strain or a further increase in the interlayer distance, especially when the graphite layers are already expanded and stabilized by defects or amorphous domains.

Motivated by this, we further explore Li and Na intercalation in the expanded AA- and AB-stacked graphite for a fixed set of interlayer distances (3.52, 3.75, 3.99, 4.25, and 4.58 Å) common for typical carbon-based electrode materials [6,30,31]. By doing so, we can effectively understand the Li and Na storage mechanisms in the expanded graphite systems, predict the critical capacity of metal ions that can be stored in the system, and address the key question of whether Na intercalation in the crystalline domains in hard carbon can actually contribute to storage capacity. The obtained energy convex hulls for Li and Na intercalation, shown in Fig. 2 (full set of the convex hulls, including both AA and AB graphite stacking, can be found in Figs. S7-S16), demonstrate that a range of concentrations have negative formation heat with respect to the pure metallic Na/Li and the expanded graphite. For the Na case, the increase in interlayer distance gradually reduces the formation heat, making the formation energy of the compound more favorable. Thus, while at 3.52 Å, no Na-C compounds are stable on the convex hull, increasing the interlayer distance to 4.58 Å for AA graphite stabilizes $NaC_{28}$, $NaC_{24}$, $NaC_{20}$, and $NaC_{16}$ compounds (see Fig. 2a, Fig. S12, and Table S5). A similar tendency is also observed for AB graphite, where Na intercalation can lead to the formation of $NaC_{28}$ compounds (also for 4.58 Å, see Fig. S12 and Table S5). If the initial interlayer distance in the electrode is close to 4.58 Å, our results thus suggest that Na can be effectively stored in graphite without a change of the interlayer distance, thus addressing, for instance, one of the main debate points in the field of hard carbon - if the Na storage can be observed within the graphite-like domains that are an inherent part of hard carbon. We also note that while the above results are presented only for the optB88-vdW exchange-correlation functional, the same tendency can also be observed for other functionals, with a small difference in the case of the maximum concentration of Na that can be realized. For instance, in DFT-D2 calculations, $NaC_6$ compounds can form (be thermodynamically stable with respect to decomposition into other products) for both 4.25 and 4.58 Å interlayer distances.

While Na and Li are very similar in terms of chemistry, due to different atomic sizes, they have some differences in the interaction chemistry with expanded graphite. Thus, for Li, the increase of the interlayer distance first increases the formation heat, while later monotonically reduces it, as shown in Fig. 2b. For instance, the computed heat of formation for $LiC_6$ for AA graphite (well-known ground-state structure in pure graphite) depends strongly on the graphite interlayer spacing: -30.20, -33.28, -26.49, -8.60, and +18.05 meV/atom at 3.52, 3.75, 3.99, 4.25, and 4.58 Å, respectively (see Table S3). This trend is expected as Li locates at hollow sites between adjacent graphene sheets and binds symmetrically to both layers. As the spacing increases from 3.52 to ≈3.75 Å, residual short-range repulsion (due to short C-Li distances) is relieved, and the formation energy becomes more favorable. The identified minimum is at 3.75 Å, which is comparable to the equilibrium interlayer distance of 3.62 Å in LiC$_6$ compound for pure AA graphite. Beyond ≈4.0 Å, the Li-graphite interaction weakens rapidly, stabilizing Li closer to one of the carbon layers, reducing the Li-C interaction energy, and ultimately rendering insertion unfavorable – at 4.58 Å interlayer distance, the heat of formation becomes positive. A similar tendency is observed for AB graphite, but with the offset of the interlayer distance at the formation heat minimum. For instance, for LiC$_{16}$, the formation heat is 6.89, -5.78, -12.02, -11.08, and -1.84 meV/atom, for spacings of 3.52, 3.75, 3.99, 4.25, and 4.58 Å, respectively. The results are not only reflected in terms of absolute values of formation heat but also in the energy convex hull. Specifically, we find that at small frozen interlayer distances (i.e., 3.52 Å, see Fig. S7), Li ions cannot intercalate into AB graphite at all - no thermodynamically stable phases are observed. In contrast, for the Li-AA graphite system, a range of compounds can be found on the energy convex hull. Moreover, for AB graphite, a significantly lower Li-content compound (maximum LiC$_{16}$ at the interlayer distance of 4.58 Å) is thermodynamically stable. The collective trends in thermodynamic stability across all concentrations and interlayer spacings are summarized by the energy above the hull for the lowest-energy structures in Fig. 2b,c. For all the Na-C and Li-C compounds, the difference in the results between AA and AB graphite can be understood in terms of registry between the two layers and the resulting asymmetry of the interlayer adsorption site. Specifically, for AA graphite, the Li or Na atom couples comparably to each carbon sheet as it locates above the hexagon ring and, thus, sees equivalent hollow sites in both adjacent layers. In AB graphite, the site is intrinsically asymmetric - the atom is bonded to 1 top C atom and a 6-member ring from the other side. These registry effects are quantitatively more pronounced for Na than for Li, which is mainly due to the larger Na atomic size and a more diffuse valence electron density; the short Na–C contact on the "near" sheet in AB graphite is significantly more repulsive at small interlayer distances than the analogous Li-C contact. Because of this, Na requires a larger interlayer spacing for favorable intercalation than Li.

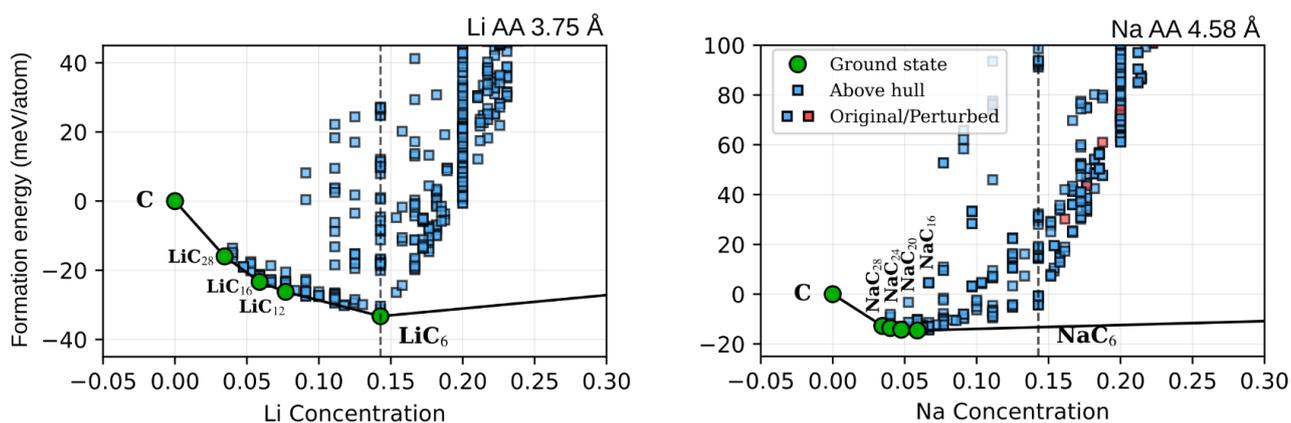
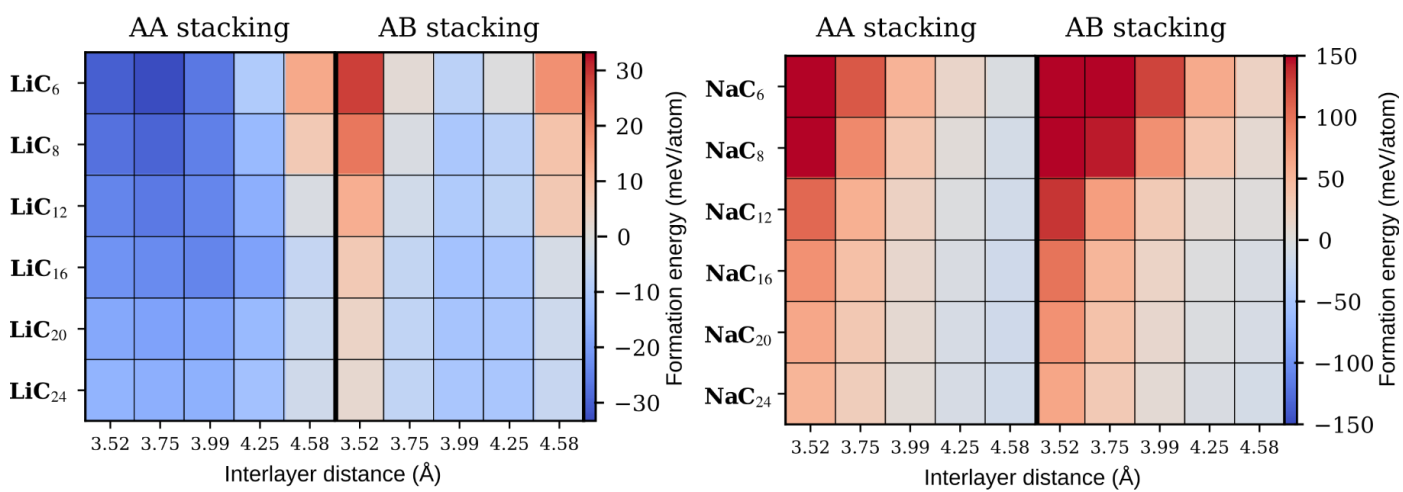
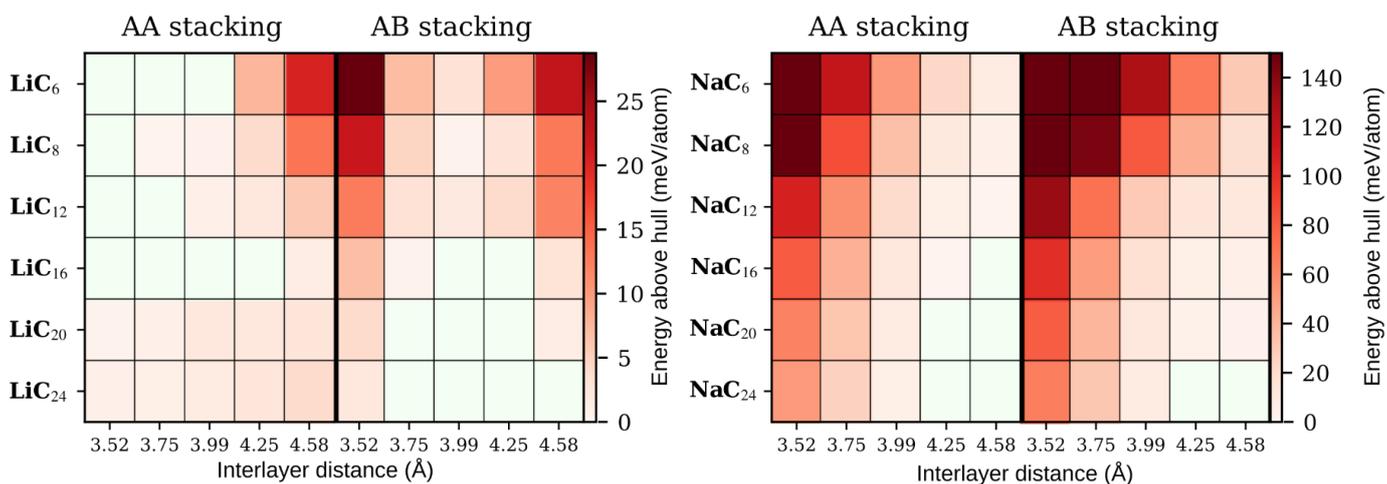

**Figure 2: Li and Na storage in expanded AA/AB graphite at a set of fixed interlayer distances calculated using the optB88-vdW functional.** (a) Representative convex hulls for Li AA (d=3.75 Å) and Na AA (d=4.58 Å) structures. The formation energies are calculated relative to pure graphite with the same stacking and d-spacing, and metallic Li/Na reference. Full data in Figs. S7-S16. (b) Changes in the formation energies with increasing d-spacing for selected Li and Na concentrations. Each datapoint corresponds to the lowest-energy structure (even if above hull) for a given Li/Na concentration, carbon stacking, and d-spacing. (c) Energy above the hull for the lowest-energy structure for a given Li/Na concentration, carbon stacking, and d-spacing.

The computed formation energies not only provide information on compounds that can be formed, but also can be used directly to predict voltage performance, as shown in Fig. 3 for different XC functionals across the range of interlayer distances. These voltage profiles correlate with the trends observed for formation energy but reflect critical insight into the electrochemical behavior of the active ions in expanded graphite. On the computational side, across all studied systems, we observe a variation in the absolute voltages depending on the choice of XC functional. Thus, similar to the formation heat calculations (Figs. S7-S16), the PBE-D2 functional predicts the highest voltages (implying the strongest interaction between extended graphite and the active ions), whereas vdW-DF and vdW-DF2 generally yield the lowest voltages (see Figs. S17-S18). For instance, for AA graphite with an interlayer distance of 3.99 Å, the voltage corresponding to the transition between $LiC_{16}$ and $LiC_6$ phases is 0.50, 0.05, 0.04, -0.15, and -0.18 V for PBE-D2, optB88-vdW, optB86b-vdW, vdW-DF2, and vdW-DF XC functionals (Fig. S17). Despite these quantitative differences, the qualitative trends regarding the effects of interlayer expansion and stacking registry are robust across all methods, with the only difference in the actual maximum storage capacity (Fig. 3b), as well as in absolute voltage values. Our results demonstrate that voltage profiles for AA graphite are significantly higher than those for AB graphite, suggesting the registry of graphite domains may change in some of the systems during Li or Na intercalation.

For Li-AA and Li-AB expanded graphite, we find that the change in the interlayer distance affects the voltage profile, with a small increase in the voltage when the interlayer distance increases from 3.52 to 3.75 Å for Li-AA and from 3.52 to 3.99 Å for Li-AB (see Fig. 3a and Fig. S19), and a reduction as the distance further increases. These results reflect the same trends observed for the heat of formation of the compounds and suggest how the interlayer distance can be used not only to maximize the concentration of Li atoms that can be stored efficiently in expanded graphite but also potentially to tune the battery voltage. For Na, the voltage profile is also sensitive to interlayer distance, but with the only difference being that, for the considered interlayer distances, monotonical increase in the voltage profile is commonly seen. Complementary to the voltage profiles, Fig. 3b highlights the importance of interlayer distance in the maximum theoretical Li/Na intercalation capacity. Specifically, for Li, the optimal interlayer spacing of 3.75 Å (for both AA-stacked graphite) supports the highest storage capacity near the commercial graphite limit of 372 mAh/g, while Li capacity drops significantly for larger interlayer distances. For instance, when the interlayer distance reaches 4.58 Å, the expected storage capacity for Li ions in AA graphite is only 80 mAh/g. These results demonstrate that high capacity for Li is tied to a specific, narrower range of interlayer distances. In contrast, for Na, the intercalation capacity (which is generally much lower than for Li, at least within the simplified model considered here) shows a direct positive correlation with d-spacing. This confirms that, for Na-ion batteries, the design goal should be to maximize the interlayer spacing as much as practically feasible, with a maximum capacity of 139 mAh/g observed at 4.58 Å for the AA-stacked graphite. The above results thus not only address the debatable questions in the field of metal-ion storage in C-based materials but also demonstrate differences in the design strategies to realize high-capacity storage of Li vs. Na in graphite-like domains in C-based electrode materials.

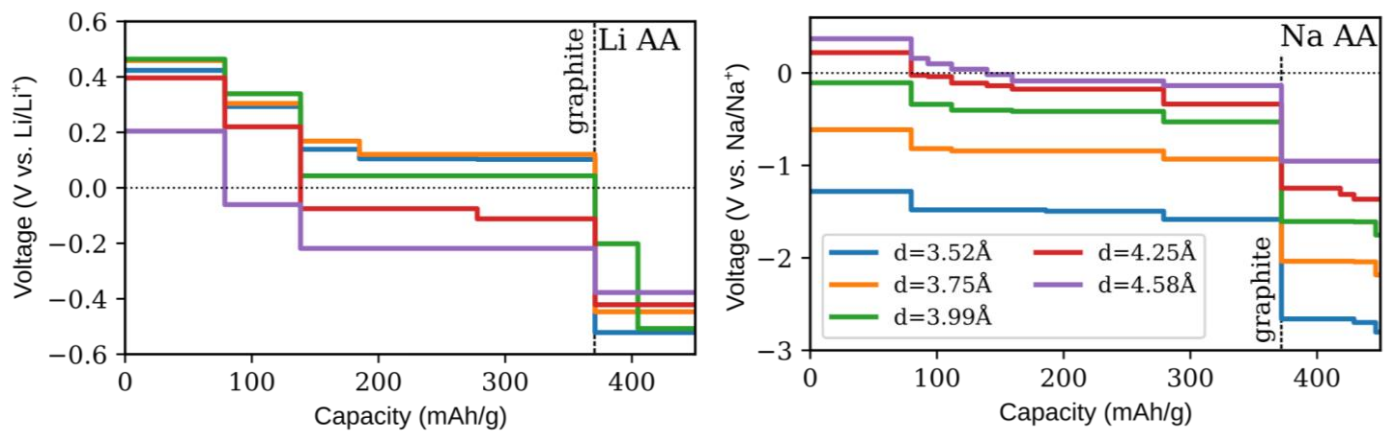

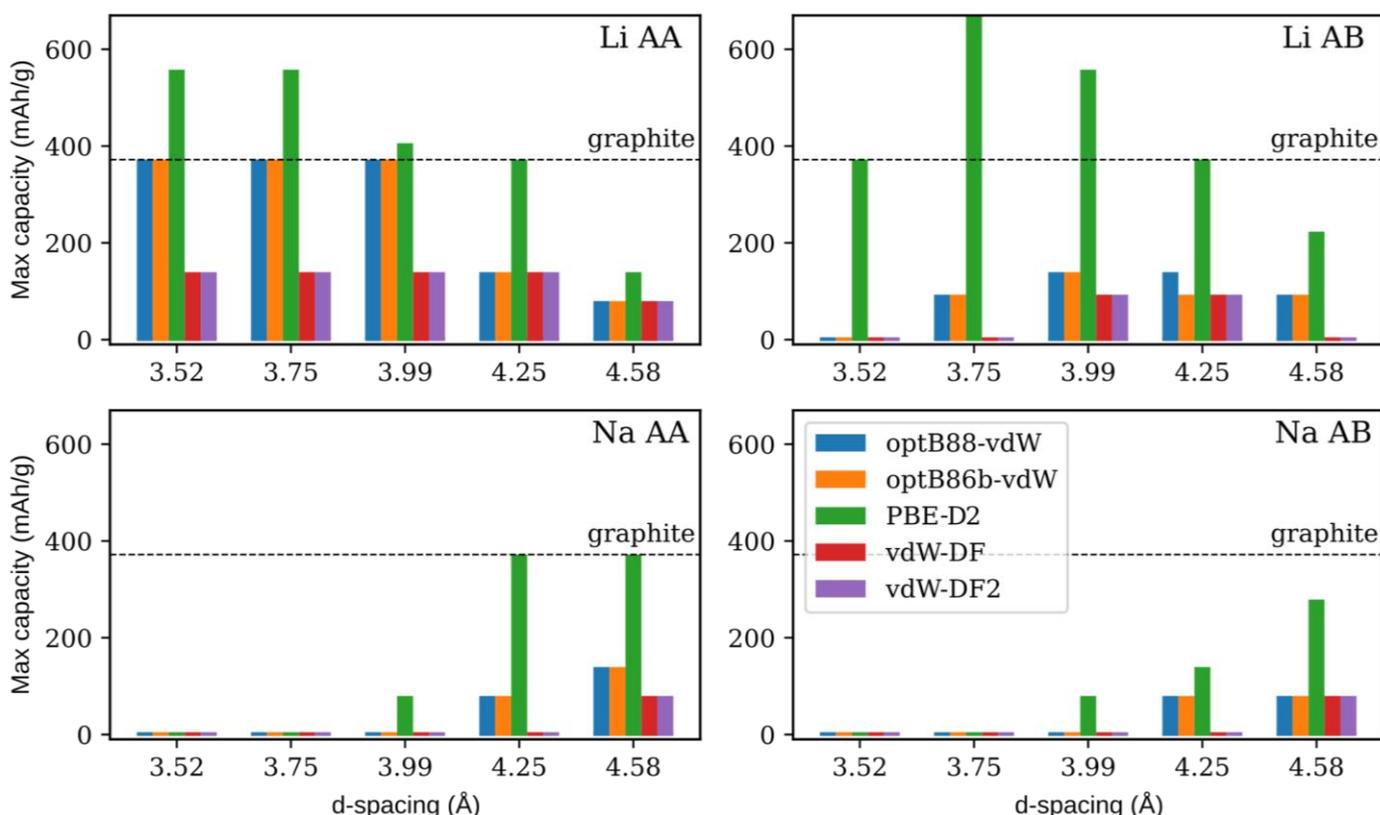

**Figure 3: Voltage profiles for Na and Li at different d-spacings and carbon stackings computed with different vdW functionals.** (a) Representative voltage profiles for Li and Na intercalation into AA-stacked carbon. The data is for the optB88-vdW functional. (b) Maximum Li/Na intercalation capacity inside AA/AB stacked carbon at different d-spacings. The data corresponds to the largest capacity exhibiting positive voltage in the corresponding voltage profiles. "Graphite" labels indicate the maximum Li capacity inside graphite (372 mAh/g).

## 3. Conclusions

By combining DFT with cluster expansion across a range of interlayer spacings and AA/AB stackings, we transform the messy structural complexity of hard carbon and expanded graphite into a clear structure-property map that yields an explicit understanding of the role of metal ion intercalation into crystalline-like domains in the electrochemical properties and establishes design principles for electrode design. Specifically, for Na-ion batteries, we demonstrate that while Na intercalation in pristine graphite is thermodynamically not possible under normal conditions, as the interlayer distance increases, Na intercalation becomes thermodynamically allowed in significant concentrations without any further change in interlayer spacing. For instance, at the interlayer spacing of 4.58 Å, expanded AA-stacked graphite can deliver 139 mAh/g (the defects/impurities are further expected to increase this capacity) within the graphite-like domains themselves. These results directly resolve a long-standing debate on the role of Na intercalation in electrochemical performance, showing that *Na can be effectively stored in crystalline domains of hard carbon/expanded graphite via intercalation reaction, provided those domains exhibit sufficiently large interlayer spacings stabilized, for instance, via impurities or an amorphous network. In such cases, the intercalation may not lead to additional XRD peak shift during cycling*. For Li, we uncover the opposite regime. Specifically, due to significantly smaller atomic size compared to Na atoms, Li storage is maximized in a narrow interlayer spacing window and rapidly decreases with the interlayer distance. At the interlayer distances of around 3.5–4.0 Å, AA graphite can support $LiC_6$ formation and capacities close to the commercial limit, while, below this window, short-range repulsion penalizes Li insertion; expanding the interlayer distance beyond ≈4.0 Å makes the Li–C interaction collapse and Li insertion progressively less favorable. These results thus demonstrate that structural motifs (and hence synthesis conditions) that are good for Na (very large interlayer spacings) are inherently harmful for Li storage inside the graphite-like domains.

## 4. Methods

All the density functional theory (DFT) calculations were done using the Vienna ab initio Simulation Package (VASP) version 6.2.1 [32–34], using the projector augmented wave (PAW) method [35]. Most of the calculations used 550 eV energy cutoffs, 10000 per reciprocal atom k-point mesh density, and a force convergence threshold of 0.01 eV/Å. Spin-polarization was only considered for pure Na and Li metals. Pure carbon, carbon-Li, and carbon-Na systems were calculated without considering spin-polarization.

To account for van der Waals (vdW) interactions in hard carbon, we screened multiple functionals – PBE-D2 [36], optB88-vdW [37], optB86b-vdW [38], vdW-DF [39], vdW-DF2 [40], and PBE-TS [41] – aiming to reproduce pure graphite d-spacing (Table S1) and Na cohesive energies (Table S4). Among the tested functionals, we found that PBE-TS yields a large difference compared to the experimentally measured Na cohesive energy (PBE-TS: -2.57 eV/atom vs. experiment: -1.13 eV/atom, see Table S4). Thus, we did not include PBE-TS into our calculations. All other functionals provide a reasonable agreement with the experiment, with optB88-vdW giving the closest graphite d-spacing, and optB86-vdW giving the closest Na cohesive energy. We decided to choose optB86-vdW as the base functional for all the calculations, later confirming our key observations by repeating a subset of the calculations using optB86b-vdW, vdW-DF, vdW-DF2, and PBE-D2 functionals.

To calculate Li and Na intercalation into expanded graphite, we used graphite structures with AA and AB stacking and fixed 3.52, 3.75, 3.99, 4.25, and 4.58 Å d-spacings. 3.52 Å is the equilibrium d-spacing for pure graphite with AA stacking calculated using the optB88-vdW functional. Other d-spacings are larger, covering the d-spacing range experimentally reported for various hard carbon structures [10]. The same d-spacings were used for graphite with both AA and AB stacking, so that the effects of stacking on Li and Na intercalation can be directly observed. In addition, we performed a set of calculations with unconstrained d-spacing, allowing the d-spacing to relax to its equilibrium state for each Na/Li-carbon system. In all these calculations, carbon atom positions were fixed, while Li/Na intercalants were allowed to relax. The lattice vectors were either fixed (for fixed d-spacing), or allowed to relax (for unconstrained d-spacing). Because AB stacking in Na/Li-graphite system tends to be unstable for non-zero Na/Li concentrations when the lattice vectors are not fixed (even if carbon atoms are not allowed to move), these Na/Li-carbon calculations were performed under additional constraint – shear stress components were fixed to prevent stacking changes while allowing d-spacing to relax.

We employed cluster expansion (CE) method (Figure 1a) as implemented in Alloy Theoretic Automated Toolkit (ATAT) version 3.36 [42,43] to find ground state Na and Li intercalation structures in AA and AB graphites with fixed 3.52 Å d-spacing.

Cluster expansion requires an *a priori* guess of the Na/Li sublattice that the corresponding metal occupies during intercalation. We assumed that Na and Li intercalants occupy the same sites inside the carbon host as the lowest-energy sites for the dilute limit (a single Na/Li atom inside a large graphite supercell, with Na/Li concentration corresponding to $MeC_{288}$ stoichiometry). To generate the sublattice, we developed a small in-house code IMDgroup-pymatgen 1.0 [44] based on pymatgen 2025.1.9 library [45] that scans the free sites inside the host carbon matrix, accounting for lattice symmetry, inserts a single Na or Li atom, and runs a series of VASP calculations to find the lowest-energy sites. The obtained Na-C and Li-C sublattices are shown on Figure S1. We tested our sublattices by performing cluster expansion with ATAT and found that the sublattices accurately describe the most stable structures at least up to $LiC_6$ and $NaC_6$ intercalant concentrations. The most stable structures for higher concentrations are sometimes better described using a different sublattice characterized by stronger Na-Na and Li-Li bonding, but we excluded these high concentrations from our analysis as their formation energies are always above the C-Li and C-Na hull.

With sublattice guess in place for Li-AA carbon, Li-AB carbon, Na-AA carbon, and Na-AB carbon (all with 3.52 Å d-spacings), we used ATAT to generate representative set of Me-C structures that is sufficient to fit the corresponding cluster expansions, find the ground state structures, and build convex hulls for each system.

Before fitting CE, we performed position relaxation for the Li/Na positions inside the generated structures. An additional self-consistent calculation was done using the tetrahedron method with Blöchl corrections without smearing to calculate accurate energies. We did not use the built-in ATAT scripts to perform the relaxation, as they are not always robust detecting issues with VASP convergence. Instead, in-house codes IMDgroup-gorun 1.0 [46] based on pymatgen 2025.1.9 [45] and ASE 3.23.0 [47] libraries were developed to (1) generate the VASP inputs from structure files generated by ATAT; (2) perform a series of VASP calculations; (3) make sure that the results are converged without issues; (4) calculate structure energy data compatible with ATAT; (5) detect and flag relaxed structures that are not compatible with the ATAT-generated structure or sublattice. It should be noted that k-point mesh generation turned out to be more robust when using kmesh command line utility distributed as a part of ATAT compared to the uniform mesh generator distributed with pymatgen 2025.1.9. Thus, we incorporated results from the kmesh utility into the in-house codes.

Energies of the relaxed structures were used to iteratively refine the CE fits. We arranged our helper codes to automatically exclude (1) structures with high, above $MeC_6$, concentrations; (2) structures were any Na/Li position, upon relaxation, moved to a different sublattice site from what is expected by ATAT, or moved further than 0.5 Å away from its sublattice site. We used ATAT to add more structures to the dataset until all the ground states with concentrations no larger than $MeC_6$ predicted by the CE fit were included and until CE cross-validation score [48] reached a sufficiently small value. At the end, all the CEs converged as follows: Li-AA carbon, d=3.52 Å – cross-validation score 1.0 meV/atom at 536 structures; Li-AB carbon, d=3.52 Å – 1.4 meV/atom at 247 structures; Na-AA carbon, d=3.52 Å – 0.5 meV/atom at 757 structures; Na-AB carbon, d=3.52 Å – 2.53 meV/atom at 188 structures. See Figures S20-S23 for more details about the CE fits.

Rather than repeating the CE fitting procedure, the obtained 4 sets of structures for the fixed 3.52 Å d-spacings were reused to find ground states for other d-spacings (3.75, 3.99, 4.25, 4.58 Å, and unconstrained). The structure sets for larger d-spacings were obtained by simple scaling of the lattice parameter perpendicular to the carbon layers. No additional CE fitting was done for these new sets of structures. Instead, the VASP calculations were repeated with the scaled structures, and new ground states were obtained. We additionally perturbed all the relaxed structures in the datasets by randomly displacing Na/Li atoms by 0.1, 0.2, and 0.4 Å, repeating the relaxation, and checking whether any of the obtained structures lay below the hull. None of the perturbation tests resulted in structures below the hull. For unconstrained d-spacings, the relaxation was implemented using ASE 3.23.0, constraining shear stress components.

Before going further, we should comment why we did not end up using the CE fitting procedure we used for the small 3.52 Å d-spacing for all the larger d-spacings. When the distance between the carbon layers with AA stacking is small, both Li and Na tend to stay in the middle between the carbon layers. However, as the d-spacing increases, the equilibrium point bifurcates, establishing two low-energy sites that are very close to each other - one is slightly closer to one carbon layer, and another slightly closer to the opposite carbon layer (see Figure S24). In this situation, the traditional cluster expansion implemented in ATAT tends to fail, generating structures with unphysically close Na-Na or Li-Li distances. More importantly, sublattice sites calculated for the dilute (or even for high $MeC_2$ concentration) limit, are no longer adequately describing the real Li and Na positions at finite concentrations – a situation that cannot be adequately described within cluster expansion formalism that assumes a fixed sublattice geometry. It is nevertheless possible to "converge" the cluster expansion by excluding structures with Li/Na positions too far from the sublattice sites (e.g. by excluding every structure with Li/Na that is more than 0.2 Å away), but comparing the results with the simple method we ended up using (with a fixed set of structures), we have found that cluster expansion misses some ground state structures despite reaching low cross-validation scores. We have done the full CE fitting for 3.99 and 4.58 Å d-spacings for all combinations of Na/Li + carbon AA/AB, and found that our methodology gave comparable results (see Figures S25-S26).

Upon calculating the ground state structures and convex hulls, we used the resulting datasets to calculate voltage profiles for Na and Li intercalation into AA and AB graphite with various d-spacings. Rather than using ground state structures with negative formation heats, we calculated the voltages using all the structures that lie on *geometric* convex hulls. The voltages were calculated as average voltages between subsequent points on a hull ($Me_{x_1}C$, $Me_{x_2}C$), following the thermodynamic formalism described by Aydinol et al [49]:

$$\bar{V} = \frac{-[G_{Me_{x_2}C} - G_{Me_{x_1}C} - (x_2 - x_1)G_{Me}]}{(x_2 - x_1)e}$$

where $G\ [eV/Me_1C]$ is the DFT-calculated energy at 0K. By using the *geometric* definition of the convex hull, we could calculate effective voltages even for structures with positive formation energy, exposing d-spacing effects on voltage even when no thermodynamically stable structures can be formed. The calculations were done using the voltage calculators available as a part of pymatgen 2025.1.9 [45].

All the earlier structure calculations were first done using the optB88-vdW functional. We also repeated all the structure relaxations and re-calculated the energies of all the reference metal structures and all the structures at least up to 50 meV/atom above the optB88-vdW-predicted hull using optB86b-vdW, PBE-D2, vdW-DF, and vdW-DF2 functionals. All the convex hulls and voltage profiles calculated using the alternative functionals can be found in the supplementary information.

## 5. Acknowledgments

The work in China was supported by the Starting Research Fund of Qingyuan Innovation Laboratory (Grant No. 00524009). The work in Poland was funded by the National Centre for Research and Development under the project WPC3/2022/50/KEYTECH/2024. Institutional and infrastructural support for the ENSEMBLE3 Centre of Excellence was provided through the ENSEMBLE3 project (MAB/2020/14) delivered within the Foundation for Polish Science International Research Agenda Programme and co-financed by the European Regional Development Fund and the Horizon 2020 Teaming for Excellence initiative (Grant Agreement No. 857543), as well as the Ministry of Education and Science initiative "Support for Centres of Excellence in Poland under Horizon 2020" (MEiN/2023/DIR/3797). The authors gratefully acknowledge access to computational resources provided by PLGrid HPC centers (ACK Cyfronet AGH, WCSS) under project number PLG/2025/018614.